\documentclass[twocolumn]{aastex61}

\received{February 4, 2017}
\revised{March 11, 2017}
\accepted{March 16, 2017}
\shorttitle{Metal mixing in dwarf galaxies}
\shortauthors{Hirai \& Saitoh}

\begin{document}

\title{Efficiency of metal mixing in dwarf galaxies}

\correspondingauthor{Yutaka Hirai}
\email{yutaka.hirai@nao.ac.jp}

\author[0000-0002-5661-033X]{Yutaka Hirai}
\affiliation{Department of Astronomy, Graduate School of Science, The University of Tokyo, 7-3-1 Hongo, Bunkyo-ku,Tokyo 113-0033, Japan}
\affiliation{Division of Theoretical Astronomy, National Astronomical Observatory of Japan, 2-21-1 Osawa, Mitaka, Tokyo 181-8588, Japan}
\affiliation{JSPS Research Fellow}
\author{Takayuki R. Saitoh}
\affiliation{Earth-Life Science Institute, Tokyo Institute of Technology, 2-12-1 Ookayama, Meguro-ku, Tokyo 152-8551, Japan}

\begin{abstract}
Metal mixing plays critical roles in the enrichment of metals in galaxies. The abundance of elements such as Mg, Fe, and Ba in metal-poor stars help us understand the metal mixing in galaxies. However, the efficiency of metal mixing in galaxies is not yet understood. Here we report a series of $N$-body/smoothed particle hydrodynamics simulations of dwarf galaxies with different efficiencies of metal mixing using turbulence-induced mixing model. We show that metal mixing apparently occurs in dwarf galaxies from Mg and Ba abundance. We find that the scaling factor for metal diffusion larger than 0.01 is necessary to reproduce the observation of Ba abundance in dwarf galaxies. This value is consistent with the value expected from turbulence theory and experiment. We also find that timescale of metal mixing is less than 40 Myr. This timescale is shorter than that of typical dynamical times of dwarf galaxies. We demonstrate that the determination of a degree of scatters of Ba abundance by the observation will help us to constrain the efficiency of metal mixing more precisely. 
\end{abstract}


\section{Introduction} \label{sec:intro}
Understanding metal mixing in galaxies gives us clues to clarify various fields in astrophysics such as the galactic chemical evolution, turbulent diffusion in the interstellar medium (ISM), and the mechanism for transition from Population III stars to Population II stars  \citep[e.g.,][]{2013RvMP...85..809K}. Metal mixing occurs in a wide range of scales from star-forming (SF) region ($\sim$ 10 pc) to large galactic scale ($\sim$ 10 kpc).

Interstellar turbulence mainly causes metal mixing. Power sources of interstellar turbulence are such as supernovae (SNe), H \textsc{ii} region, fluid instabilities, galactic rotation, and galactic interactions \citep[e.g.,][]{2004ARA&A..42..211E}. Observations of power spectra of CO emission, H \textsc{i} emission, and absorption can be used to determine power law slopes of the spectra that characterizes the interstellar turbulence.

Elemental abundance ratios in metal-poor stars give us signatures to constrain metal mixing process. Astronomical spectroscopic observations of extremely metal-poor (EMP) stars in the Local Group galaxies have shown that star-to-star scatters of $\alpha$-elements are less than 0.2 dex, which is comparable to the observational errors \citep[e.g.,][]{2015ARA&A..53..631F}. 

On the other hand, elements synthesized by rapid neutron capture process ($r$-process elements) show scatters of over 3 dex in EMP stars in the Milky Way (MW) halo \citep[see Figure 10 in][]{2015ARA&A..53..631F}. The scatters of $r$-process elements decrease as the metallicity increases. The Local Group dwarf spheroidal galaxies (dSphs) show lower abundances of $r$-process elements than that of the Milky Way halo. There is no very metal-poor stars with [Ba/Fe]\footnote{[A/B] = $\log_{10}({N_{\mathrm{A}}}/{N_{\mathrm{B}}})-\log_{10}({N_{\mathrm{A}}}/{N_{\mathrm{B}}})_{\odot}$, where $N_{\mathrm{A}}$ and $N_{\mathrm{B}}$ are number densities of elements A and B, respectively.} $\gtrsim$ 1 in dSphs \citep{2015ARA&A..53..631F}. Only Reticulum II ultrafaint dwarf galaxy has enhanced $r$-process abundance \citep{2016Natur.531..610J, 2016AJ....151...82R}. These observational facts would be a piece of evidence to constrain metal mixing in galaxies.

The different behavior of the abundance of $\alpha$- and $r$-process elements reflects the astrophysical site of these elements. In the case of core-collapse supernovae (CCSNe), they synthesize both $\alpha$-elements and iron \citep[e.g.,][]{2013ARA&A..51..457N}. This means that inhomogeneity in the abundance of Fe and Mg is related in many cases, i.e., if Fe is high in a region, Mg is also high. The scatters of [$\alpha$/Fe] in EMP stars are thus mainly determined by the yields of CCSNe. The scatters in the ratio of $\alpha$-elements and iron are therefore only $\sim$ 0.2 dex. On the other hand, although the astrophysical site of $r$-process elements is not yet well understood, neutron star mergers (NSMs) are one of the most likely sites of $r$-process elements \citep[e.g.,][]{2014ApJ...789L..39W}. In this case, $r$-process elements and iron are produced in a different site. This means that the ratio of $r$-process elements and iron easily reflect the inhomogeneity of these elements \citep{2017MNRAS.466.2474H}. Lower abundance of $r$-process elements in dSphs suggests that the metal mixing was efficient in these galaxies.

Previous hydrodynamic simulations have shown that metal mixing significantly affects the metal content in galaxies. \citet{2010MNRAS.407.1581S} introduced a metal diffusion model in their smoothed particle hydrodynamic (SPH) simulations following \citet{1963MWRv...91...99S}. Their results suggest that metal diffusion increases the metal content of gas in galaxies and decrease it in the intergalactic medium. \citet{2016A&A...588A..21R} systematically studied the scatters of $\alpha$-elements with different parameters. They show that metal mixing is required to reduce the scatters of $\alpha$-elements in particle-based simulations. \citet{2016ApJ...822...91W} studied effects of different metal diffusion strength. They found that a stronger diffusion produces a tighter [O/Fe] versus [Fe/H]. They also found that efficiency of metal mixing does not strongly affect stellar metallicity distribution.

However, the efficiency of metal mixing is not yet understood. \citet{2016A&A...588A..21R} pointed out that the scaling factor for their metal diffusion of 10$^{-3}$ reproduce the low scatters of $\alpha$-elements in dSphs, but more efficient mixing is required to reproduce that of a Milky Way-like galaxy. Low scatters of $\alpha$-elements ($\sim$ 0.2 dex) make it difficult to constrain the efficiency of metal mixing. This value is comparable to the observational errors. We need additional signatures to constrain the efficiency of metal mixing.

The aim of this paper is to constrain the metal mixing through a series of hydrodynamic simulations of galaxies and abundances of $r$- and $\alpha$-elements. We study metal mixing in dSphs. They are the ideal site to study metal mixing because of their simple structures \citep[e.g.,][]{2013ApJ...774..103L, 2015ApJ...807..154B}. In this paper, we examine how the metal mixing affects the chemical abundance in dSphs.
\section{Method} \label{sec:method}
\subsection{Code}
Here, we describe the methods adopted in this study. \citet{2015ApJ...814...41H} and \citet{2017MNRAS.466.2474H} show details of our implementation. We perform a series of $N$-body/SPH simulations using \textsc{asura} \citep{2008PASJ...60..667S, 2009PASJ...61..481S}. Hydrodynamics are computed using density-independent SPH (DISPH) method \citep{2013ApJ...768...44S}. The DISPH method enables us to treat the contact discontinuity and fluid instability correctly. It can treat metal mixing more accurate than standard SPH method. To accelerate the integration of self-gravitating fluid systems, we use a Fully Asynchronous Split Time-Integrator for a Self-gravitating Fluid (FAST) scheme \citep{2010PASJ...62..301S}. Time step limiter is also implemented to handle the strong shock correctly \citep{2009ApJ...697L..99S}. 

For the calculation of cooling and heating, we adopt the metallicity dependent cooling/heating function from 10 to 10$^9$ K generated by \textsc{cloudy} \citep{1998PASP..110..761F, 2013RMxAA..49..137F}. We implement heating from the cosmic ultraviolet background radiation following \citet{2012ApJ...746..125H}. We also implement the effects of self-shielding in the dense interstellar gas following the fitting function presented in \citet{2013MNRAS.430.2427R}.

Star formation criteria are based on \citet{2008PASJ...60..667S, 2009PASJ...61..481S}. When a gas particle satisfies three conditions: converging flow ($\nabla\cdot\mbox{\boldmath{v}}<$ 0, $\mbox{\boldmath{v}}$: velocity of gas), high number density ($>$ 100 cm$^{-3}$), and low temperature ($T~<$ 1000 K), it is converted into a star particle.

We treat each star particle as a single stellar population (SSP). We adopt the initial mass function of \citet{2003PASP..115..763C} from 0.1 to 100 $M_{\sun}$. Thermal and chemical feedback from SSP particles is computed using Chemical Evolution Library \citep[\textsc{celib},][]{2016ascl.soft12016S,2017AJ....153...85S}. We adopt the reference feedback models in \textsc{celib}. We assume that stars more massive than 8 $M_{\sun}$ stochastically explode as CCSNe and distribute the thermal energy of $10^{51}$ erg to the surrounding gas particles \citep{2008MNRAS.385..161O}. For the yields of CCSNe, we adopt that of \citet{2013ARA&A..51..457N}. We assume that 5 \% of stars with more massive than 20 $M_{\sun}$ explode as hypernovae based on observations of gamma-ray bursts \citep{2007ApJ...657L..73G}.

We have also adopted models of type Ia supernovae (SNe Ia) in \textsc{celib}. We adopt the empirical delay time distribution of SNe Ia with the power law index of $-1$ \citep{2012PASA...29..447M}. We set the minimum delay time of SNe Ia to be 10$^8$ yr \citep{2008PASJ...60.1327T}. The energy from SNe Ia is distributed in the same way as CCSNe. We adopt the yields of SNe Ia from the model N100 of \citet{2013MNRAS.429.1156S}. Here we adopt Mg as a representative of $\alpha$-elements because there are large enough observational data to compare with our simulation.

In this study, we assume that NSMs synthesize $r$-process elements. We implement the merger time distribution of NSMs in a similar way as SNe Ia. According to population synthesis calculations in \citet{2012ApJ...759...52D}, the merger time distribution is close to the power law with its index of $-1$. We set minimum merger time as 10$^7$ yr following \citet{2012ApJ...759...52D}. We assume that 1 \% of stars with 20 to 40 $M_{\sun}$ cause NSMs following \citet{2015ApJ...814...41H}. This rate corresponds to 50 Myr$^{-1}$ in the MW mass galaxies, which is consistent with the estimation using observed binary pulsars \citep{2008LRR....11....8L}. We distribute the ejecta of $r$-process elements with the same way as SNe for simplicity. Ejecta from NSMs might expand larger region than SNe \citep{2014A&A...565L...5T}. We discuss this effect in Section \ref{Discussion}. For yields of NSMs, we assume that each NSM ejects 1.8 $\times$ 10$^{-4}$ $M_{\sun}$ of Ba following \citet{2015ApJ...804L..35I}. Although $s$-process synthesizes 85 \% of Ba, we follow Ba as $r$-process elements. We can eliminate $s$-process contribution by using the [Ba/Eu] ratio because $r$-process synthesizes 99\% of Eu \citep{2000ApJ...544..302B}. The pure $r$-process [Ba/Eu] ratio is [Ba/Eu] = $-$0.89 \citep{2000ApJ...544..302B}. The contribution of $s$-process mostly begins from [Fe/H] = $-$2.75 in the MW halo \citep{2000ApJ...544..302B}. On the other hand, the contribution of $s$-process would appear lower metallicity in dSphs. However, the number of stars that Eu has been detected is small. We, therefore, correct the observational value of [Ba/Eu] to be $-$0.89 for all stars with [Fe/H] $>-$2.75.

%Diffusion
For metal diffusion, we have implemented turbulent metal mixing model in DISPH \citep{2017AJ....153...85S} based on \citet{1963MWRv...91...99S} and \citet{2010MNRAS.407.1581S}. The $i$th metal, $Z_{i}$, diffuses to surrounding gas particles with the following equation,
\begin{eqnarray}\label{Diff}
\frac{{\rm{d}}Z_i}{{\rm{d}}t} &=&  \nabla(D{\nabla}Z_i),\nonumber\\
D &=& C_{\rm{d}}|S_{ij}|h^{2},
\end{eqnarray}
where $C_{\rm{d}}$ is the scaling factor for metal diffusion, $S_{ij}$ is the trace-free shear tensor, and $h$ is the smoothing length of SPH. \citet{2010MNRAS.407.1581S} show details of the implementation. 
\subsection{Isolated dwarf galaxy model}
We adopt the isolated dwarf galaxy model presented in \citet{2012A&A...538A..82R} and \citet{2017MNRAS.466.2474H}. We confirm that this model can reproduce observed properties, such as metallicity distribution function, mass--metallicity relation, and velocity dispersion profiles, of the Local Group dwarf galaxies \citep{2015ApJ...814...41H, 2017MNRAS.466.2474H}. The total number of particles is $2^{18}$. The halo mass and central density are $7\times10^{8}~M_{\sun}$ and $5.0 \times 10^7 M_{\sun}$ kpc$^{-3}$, respectively. Mass of one dark matter and gas particles are 4.5$\times$10$^3~M_{\sun}$ and 8.0$\times$10$^{2}$ $M_{\sun}$, respectively. We have confirmed that effects of IMF sampling \citep{2016A&A...588A..21R} do not strongly affect our results. We have also confirmed that changing the resolution of the simulation does not strongly affect the results. We set gravitational softening length as 7.8 pc. Final stellar mass of this galaxy is $5\times10^{6}~M_{\sun}$. In this study, we varied the value of $C_{\rm{d}}$ with a fixed galaxy model. Table \ref{models} lists adopted value of $C_{\rm{d}}$.

\begin{deluxetable}{ll}
   \tabletypesize{\scriptsize}
   \tablecaption{Adopted value of scaling factor for metal diffusion. \label{models}} 
   \tablecolumns{2}
\tablenum{1}
   \tablewidth{0pt}
   \tablehead{
      \colhead{Model}& 
      \colhead{$C_{\rm d}$}}
      \startdata
      d1000&0.1\\
      d0100&0.01\\
      d0010&0.001\\
      d0001&0.0001\\
      d0000&\nodata \\
      \enddata
     \tablecomments{First column shows the name of models. Second column shows the value of scaling factor for metal diffusion.}
\end{deluxetable}

\section{Results} \label{sec:results}
\begin{figure*}[htbp]
\epsscale{1.0}
\plotone{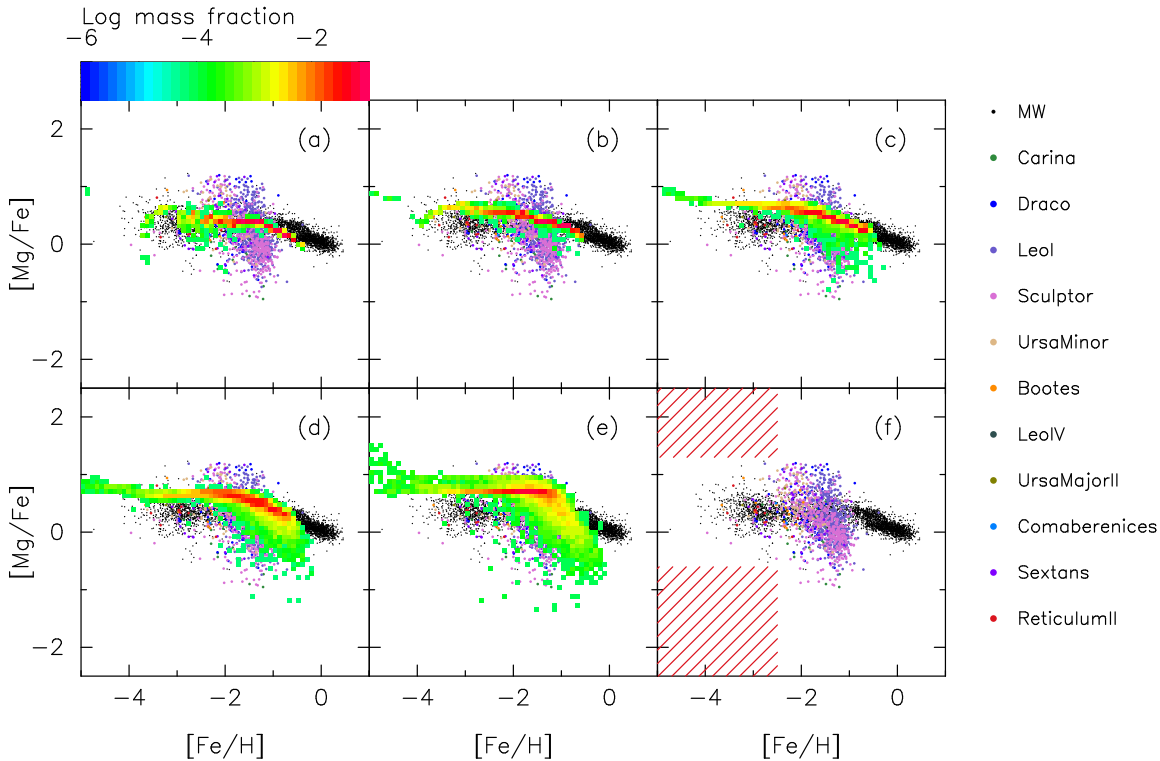}
\caption{Stellar [Mg/Fe] as a function of [Fe/H]. Panels (a), (b), (c), (d), and (e) represent results of models d1000, d0100, d0010, d0001, and d0000, respectively. Panel (f) represents observational data. Red hatched area shows the region that there are no stars with Mg. Color contours show computed stellar abundances in the logarithm of a mass fraction in each grid. Black dots represent the observed abundances of the Milky Way. Colored dots represent the observed abundances of the Local Group dwarf galaxies (green: Carina, blue: Draco, purple: LeoI, magenta: Sculptor, ocher: Ursa Minor, orange: Bootes, dark green: LeoIV, grass green: Ursa Major II, sky-blue: Coma Berenices, red-purple: Sextans, and red: Reticulum II). We compile all observed data using SAGA database \citep{2008PASJ...60.1159S, 2011MNRAS.412..843S, 2014MmSAI..85..600S, 2013MNRAS.436.1362Y}. \label{MgFe}}
\end{figure*}

Figure \ref{MgFe} shows [Mg/Fe] as a function of [Fe/H]. Computed scatters of [Mg/Fe] lie within the observed scatters at [Fe/H] $<-$2 when we implement the metal mixing (Figure \ref{MgFe}a--d). On the other hand, model d0000 has several stars with [Mg/Fe] $>1$ or [Mg/Fe] $<-1$ at [Fe/H] $\lesssim-2.5$ (red hatched area in Figure \ref{MgFe}f). These stars have not reported in the observation. In the SPH scheme, the abundance of metals in each SPH particle cannot diffuse to the surrounding gas particles if we do not implement metal mixing. Model d0000 forms these stars with an artificial effect. These results indicate that we need to implement the metal mixing model in SPH simulations when we discuss the scatters of elemental abundance ratios.

However, it is difficult to constrain the efficiency of metal mixing only through the [Mg/Fe] ratio. All models with metal mixing show the smaller dispersion of [Mg/Fe] than that of the observation at [Fe/H] $<-$2. Models d0010 and d0001 seem to reproduce the sub-solar [Mg/Fe] ratio seen in [Fe/H] $>-2$, but the statistical and observational errors of these stars typically have over 0.5 dex. We cannot conclude that models d0010 and d0001 are superior to models d1000 and d0100. 

%----[Mg/Fe] as a function of [Fe/H]-----------------------------------------------------

\begin{figure*}[htbp]
\epsscale{1.0}
\plotone{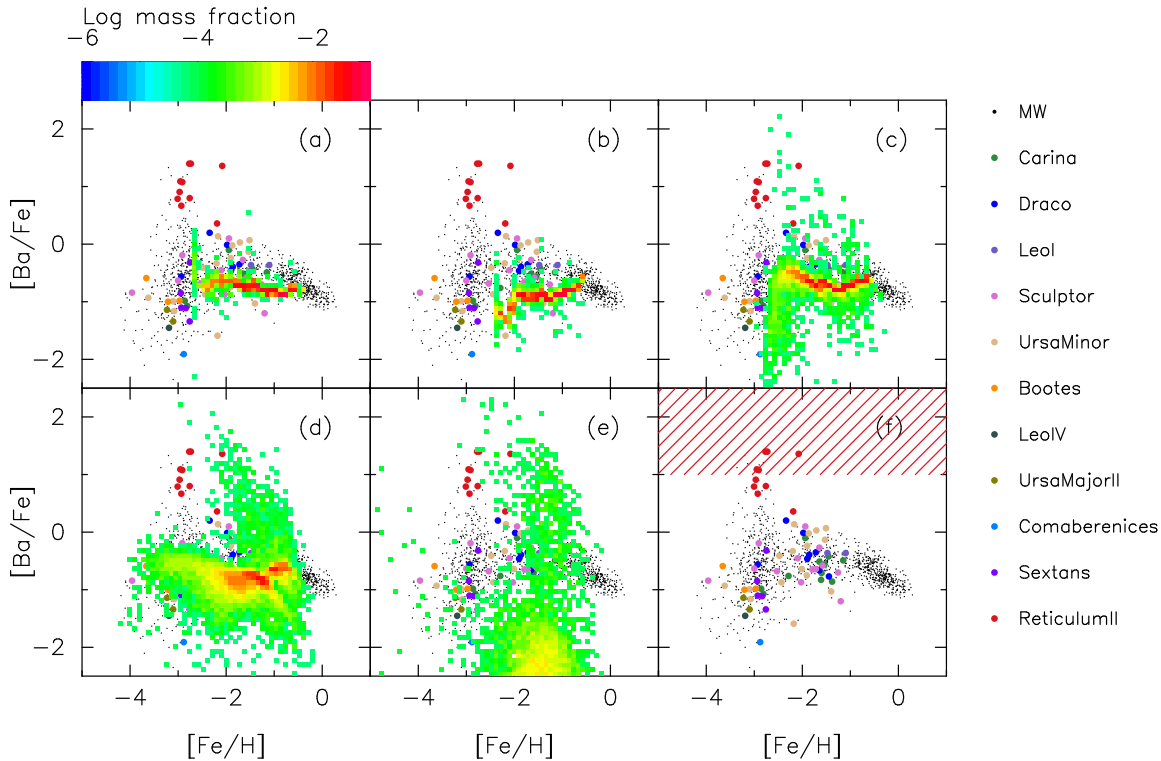}
\caption{Stellar [Ba/Fe] as a function of [Fe/H]. Panels (a), (b), (c), (d), and (e) represent results of models d1000, d0100, d0010, d0001, and d0000, respectively. Panel (f) represents observational data. Red hatched area shows region that there is no stars with Ba in dSphs except for Reticulum II ultrafaint dwarf. Symbols are the same as in Figure \ref{MgFe}. We correct the observational value of [Ba/Eu] to be $-$0.89 for all stars with [Fe/H] $>-$2.75. We compile all observed data using SAGA database \citep{2008PASJ...60.1159S, 2011MNRAS.412..843S, 2014MmSAI..85..600S, 2013MNRAS.436.1362Y}.\label{BaFe}}
\end{figure*}

In the case of [Ba/Fe], the metal mixing is expected to be more important than [Mg/Fe] because [Ba/Fe] shows large scatters in [Fe/H] $\lesssim -$3. Figure \ref{BaFe} shows [Ba/Fe] as a function of [Fe/H]. As shown in Figure \ref{BaFe}, the difference of the efficiency of metal mixing alters the scatters of [Ba/Fe] significantly. Models with $C_{\rm d} \leq$ 0.001 produce stars with [Ba/Fe] $>$ 1 (Figure \ref{BaFe} c--e). These stars are located on the red hatched area in Figure \ref{BaFe}f, i.e., not reported in the observation. On the other hand, models d1000 and d0100 do not have stars with [Ba/Fe] $>$1 due to efficient metal mixing. This result is consistent with the observation which shows no stars with [Ba/Fe] $>$ 1 at [Fe/H] $<-$2.5.

%add the revision of (1)
The value of $C_{\rm d}$ controls the fraction of stars which highly deviate from the average value of metallicity. Models with the higher value of $C_{\rm d}$ form a larger fraction of stars around average values of metallicity in each epoch. This effect can be seen in Figure \ref{BaFe}. Ba abundance first appears at [Fe/H] $\sim$ $-$3 in models with high values of $C_{\rm d}$ (Figure \ref{BaFe}a, b, and c). This metallicity reflects the average value of [Fe/H] at the time first NSM occurs. Note that the different ranges of [Fe/H] between Figures \ref{MgFe} and \ref{BaFe} reflect the delayed production of Ba due to long merger times of NSMs.

%----[Ba/Fe] as a function of [Fe/H]-----------------------------------------------------

\section{Discussion}\label{Discussion}
In the last section, we show that the value of $C_{\rm d}$ $\gtrsim$ 0.01 is necessary to prevent forming stars with [Ba/Fe] $\gtrsim$ 1 in dSphs. This result suggests that the formation of these stars, which does not exist in the Local Group dSphs, is strongly related to the efficiency of metal mixing. When NSMs occur in a galaxy, some surrounding gas will have the value of [Ba/Fe] $\gtrsim$ 1. This gas mixes into gas with lower $r$-process abundance. Stars with high $r$-process abundance are only formed when gas with high $r$-process abundance remains in a galaxy. This means that the timescale of metal mixing has the critical role in forming these stars.

Here we compute the average time that the gas with [Ba/Fe] $>$ 1 becomes less than [Ba/Fe] $<$ 0 (hereafter we define this time as $t_{\rm mix}$). We find that models d1000 and d0100 have $t_{\rm mix}$ $\lesssim$ 10 and $\simeq$ 40 Myr, respectively\footnote{We cannot compute the time less than 10 Myr because the time interval of the snapshots in this simulation is 10 Myr.}. These values are smaller than the dynamical time of the system ($\sim$ 100 Myr), which determines an early star formation rate in a galaxy \citep{2017MNRAS.466.2474H}. On the other hand, models d0010, and d0001 have $t_{\rm mix}$ $\simeq$ 360 Myr and 1.6 Gyr, respectively. These results suggest that the $r$-process elements are mixed with shorter timescale than the timescale of SF in galaxies.

Stars with high $r$-process abundance are only formed in ultrafaint dwarf galaxies with initial gas mass with $\sim$ 10$^6$ $M_{\sun}$ \citep{2016Natur.531..610J}. In these galaxies, even if all metals are mixed in a whole galaxy, gas mass is not enough to prevent forming $r$-process enhanced stars. These results imply that $r$-process enhanced stars in the MW halo are originated from ultrafaint dwarf galaxies accreted to the MW halo.

%add the revision of (2)
In this letter, we adopt the isolated dwarf galaxy model. According to cosmological hydrodynamic simulations of  \citet{2017arXiv170207355J}, lowest metallicity stars are formed in external enriched halos accreted later in the main halo. If we consider a fully cosmological structure formation, lower metallicity stars formed by external enrichment would be seen such as in Figure \ref{BaFe}a, b, and c. We expect that these external enriched stars have low [Ba/Fe] value as well as [Fe/H] value because both Fe and Ba would be well diluted in ISM when they reach the subhalo around the main halo. These stars may account for observed stars with [Fe/H] $< -$3.5 and [Ba/Fe] $<$ 0.0 plotted in Figure \ref{BaFe}.

The diffusion coefficient inferred from the Ba abundance is consistent with the theoretical and experimental value. If we assume a Kolmogorov constant as 1.41, the value of $C_{\rm d}$ is 0.0324 \citep[e.g., ][]{2000AnRFM..32....1M}. In the case of turbulent mixing layer and turbulent channel flow, optimal values of $C_{\rm d}$ are 0.0225 and 0.01, respectively \citep{1987JCoPh..71..343H}. Although the Smagorinsky model has a problem that the coefficient may depend on the flow field, our results can be an independent experiment of the Smagorinsky model on the point of metal mixing in galaxies.

%add the revision of (3)
The metal mixing efficiency of models with $C_{\rm d}$ $\gtrsim$ 0.01 also consistent with models of \citet{2015MNRAS.454..659J}.  They find that signatures from Population III stars efficiently wiped out by Population II supernovae. Their adopted diffusion coefficients are 2.4 $\times$ 10$^{-5}$ kpc$^2$ Myr$^{-1}$ and 8.1 $\times$ 10$^{-4}$ kpc$^2$ Myr$^{-1}$ for their minihalo and atomic cooling halo models, respectively. The average diffusion coefficients calculated following equation (\ref{Diff}) at 1 Gyr from the beginning of our simulation are 2 $\times$ 10$^{-4}$ kpc$^2$ Myr$^{-1}$ and 2 $\times$ 10$^{-5}$ kpc$^2$ Myr$^{-1}$ for models d1000 and d0100, respectively. This result means that we can draw similar conclusion with \citet{2015MNRAS.454..659J} if we adopt the metal mixing parameter of $C_{\rm d}$ $\gtrsim$ 0.01.

In this study, we adopt the metal diffusion model of \citet{2010MNRAS.407.1581S}. \citet{2009MNRAS.392.1381G} also constructed metal diffusion model. They used velocity dispersion to determine the diffusion coefficient. \citet{2016ApJ...822...91W} compared difference between diffusion model of \citet{2009MNRAS.392.1381G} and \citet{2010MNRAS.407.1581S}. They found that diffusion coefficient computed in \citet{2009MNRAS.392.1381G} is about twice as large as that of \citet{2010MNRAS.407.1581S}. They also found that the gas properties do not depend on the method to calculate the diffusion coefficient. We confirm that adopting diffusion model of \citet{2009MNRAS.392.1381G} show almost the same results with the results presented in this paper.

In this letter, we treat ejecta of SNe and NSMs with the same way. On the other hand, the ejecta from NSMs would expand larger region than that of SNe because NSMs eject materials with higher speed ($\sim$ 0.1$c$) than that of SNe \citep{2014A&A...565L...5T}. \citet{2017ApJ...835L...3T} estimated that the mass of interstellar materials mixed by ejecta from NSMs is $\sim$ 3.5$\times~10^{6}~M_{\sun}$, which is 100 times larger than that of SNe. In this case, fewer stars with an extremely high abundance of $r$-process elements would be formed. However, if Fe is not efficiently mixed, many stars with [Ba/Fe] $<$ 0 and [Fe/H] $>-$2, which are not observed, are still formed. This means that the difference of the ejection mechanism between NSMs and SNe does not strongly alter the results.

To more precisely constrain the efficiency of metal mixing, we need to observe a greater number of metal-poor stars in dSphs. It is important to determine the degree of scatters in the abundance of $r$-process elements to constrain the efficiency of metal mixing. Besides, we find that the fraction of EMP stars is smaller in models with more efficient metal mixing. Although several EMP stars are recently reported in ultrafaint dwarf galaxies, \citet{2006ApJ...651L.121H} reported that the lack of EMP stars in dSphs. This result would be related to the efficiency of metal mixing. We expect that fraction of EMP stars in each dSph would also be an indicator of the efficiency of metal mixing.

\acknowledgements
We are grateful for the anonymous referee for giving us helpful comments. This work was supported by JSPS KAKENHI Grant Numbers 15J00548, 26707007, MEXT SPIRE, and JSPS and CNRS under the Japan-France Research Cooperative Program. Numerical computations and analysis were in part carried out on Cray XC30 and computers at CfCA, NAOJ. This research has made use of NASA's Astrophysics Data System.

%\bibliography{sampleNotes}

\begin{thebibliography}{}
\bibitem[Bland-Hawthorn et al.(2015)]{2015ApJ...807..154B} Bland-Hawthorn, J., Sutherland, R., \& Webster, D.\ 2015, \apj, 807, 154 
\bibitem[Burris et al.(2000)]{2000ApJ...544..302B} Burris, D.~L., Pilachowski, C.~A., Armandroff, T.~E., et al.\ 2000, \apj, 544, 302
\bibitem[Chabrier(2003)]{2003PASP..115..763C} Chabrier, G.\ 2003, \pasp, 115, 763
\bibitem[Dominik et al.(2012)]{2012ApJ...759...52D} Dominik, M., Belczynski, K., Fryer, C., et al.\ 2012, \apj, 759, 52
\bibitem[Elmegreen \& Scalo(2004)]{2004ARA&A..42..211E} Elmegreen, B.~G., \& Scalo, J.\ 2004, \araa, 42, 211
\bibitem[Ferland et al.(1998)]{1998PASP..110..761F} Ferland, G.~J., Korista, K.~T., Verner, D.~A., et al.\ 1998, \pasp, 110, 761 
\bibitem[Ferland et al.(2013)]{2013RMxAA..49..137F} Ferland, G.~J., Porter, R.~L., van Hoof, P.~A.~M., et al.\ 2013, \rmxaa, 49, 137 
\bibitem[Frebel \& Norris(2015)]{2015ARA&A..53..631F} Frebel, A., \& Norris, J.~E.\ 2015, \araa, 53, 631
\bibitem[Greif et al.(2009)]{2009MNRAS.392.1381G} Greif, T.~H., Glover, S.~C.~O., Bromm, V., \& Klessen, R.~S.\ 2009, \mnras, 392, 1381
\bibitem[Guetta \& Della Valle(2007)]{2007ApJ...657L..73G} Guetta, D., \& Della Valle, M.\ 2007, \apjl, 657, L73 
\bibitem[Haardt \& Madau(2012)]{2012ApJ...746..125H} Haardt, F., \& Madau, P.\ 2012, \apj, 746, 125
\bibitem[Helmi et al.(2006)]{2006ApJ...651L.121H} Helmi, A., Irwin, M.~J., Tolstoy, E., et al.\ 2006, \apjl, 651, L121 
\bibitem[Hirai et al.(2015)]{2015ApJ...814...41H} Hirai, Y., Ishimaru, Y., Saitoh, T.~R., et al.\ 2015, \apj, 814, 41
\bibitem[Hirai et al.(2017)]{2017MNRAS.466.2474H} Hirai, Y., Ishimaru, Y., Saitoh, T.~R., et al.\ 2017, \mnras, 466, 2474
\bibitem[Horiuti(1987)]{1987JCoPh..71..343H} Horiuti, K. \ 1987, Journal of Computational Physics, 71, 343
\bibitem[Ishimaru et al.(2015)]{2015ApJ...804L..35I} Ishimaru, Y., Wanajo, S., \& Prantzos, N.\ 2015, \apjl, 804, L35
\bibitem[Ji et al.(2015)]{2015MNRAS.454..659J} Ji, A.~P., Frebel, A., \& Bromm, V.\ 2015, \mnras, 454, 659  
\bibitem[Ji et al.(2016)]{2016Natur.531..610J} Ji, A.~P., Frebel, A., Chiti, A., \& Simon, J.~D.\ 2016, \nat, 531, 610
\bibitem[Jeon et al.(2017)]{2017arXiv170207355J} Jeon, M., Besla, G., \& Bromm, V.\ 2017, arXiv:1702.07355 
\bibitem[Karlsson et al.(2013)]{2013RvMP...85..809K} Karlsson, T., Bromm, V., \& Bland-Hawthorn, J.\ 2013, Reviews of Modern Physics, 85, 809 
\bibitem[Lee et al.(2013)]{2013ApJ...774..103L} Lee, D.~M., Johnston, K.~V., Tumlinson, J., Sen, B., \& Simon, J.~D.\ 2013, \apj, 774, 103 
\bibitem[Lorimer(2008)]{2008LRR....11....8L} Lorimer, D.~R.\ 2008, Living Reviews in Relativity, 11, 8
\bibitem[Maoz \& Mannucci(2012)]{2012PASA...29..447M} Maoz, D., \& Mannucci, F.\ 2012, \pasa, 29, 447
\bibitem[Meneveau \& Katz(2000)]{2000AnRFM..32....1M} Meneveau, C., \& Katz, J.\ 2000, Annual Review of Fluid Mechanics, 32, 1
\bibitem[Nomoto et al.(2013)]{2013ARA&A..51..457N} Nomoto, K., Kobayashi, C., \& Tominaga, N.\ 2013, \araa, 51, 457
\bibitem[Okamoto et al.(2008)]{2008MNRAS.385..161O} Okamoto, T., Nemmen, R.~S., \& Bower, R.~G.\ 2008, \mnras, 385, 161 
\bibitem[Rahmati et al.(2013)]{2013MNRAS.430.2427R} Rahmati, A., Pawlik, A.~H., {Rai{\v c}evi{\` c}}, M., \& Schaye, J.\ 2013, \mnras, 430, 2427
\bibitem[Revaz et al.(2016)]{2016A&A...588A..21R} Revaz, Y., Arnaudon, A., Nichols, M., Bonvin, V., \& Jablonka, P.\ 2016, \aap, 588, A21
\bibitem[Revaz \& Jablonka(2012)]{2012A&A...538A..82R} Revaz, Y., \& Jablonka, P.\ 2012, \aap, 538, A82
\bibitem[Roederer et al.(2016)]{2016AJ....151...82R} Roederer, I.~U., Mateo, M., Bailey, J.~I., III, et al.\ 2016, \aj, 151, 82 
\bibitem[Saitoh(2016)]{2016ascl.soft12016S} Saitoh, T.~R.\ 2016, Astrophysics Source Code Library, ascl:1612.016 
\bibitem[Saitoh(2017)]{2017AJ....153...85S} Saitoh, T.~R.\ 2017, \aj, 153, 85
\bibitem[Saitoh et al.(2008)]{2008PASJ...60..667S} Saitoh, T.~R., Daisaka, H., Kokubo, E., et al.\ 2008, \pasj, 60, 667 
\bibitem[Saitoh et al.(2009)]{2009PASJ...61..481S} Saitoh, T.~R., Daisaka, H., Kokubo, E., et al.\ 2009, \pasj, 61, 481
\bibitem[Saitoh \& Makino(2009)]{2009ApJ...697L..99S} Saitoh, T.~R., \& Makino, J.\ 2009, \apjl, 697, L99
\bibitem[Saitoh \& Makino(2010)]{2010PASJ...62..301S} Saitoh, T.~R., \& Makino, J.\ 2010, \pasj, 62, 301
\bibitem[Saitoh \& Makino(2013)]{2013ApJ...768...44S} Saitoh, T.~R., \& Makino, J.\ 2013, \apj, 768, 44
\bibitem[Seitenzahl et al.(2013)]{2013MNRAS.429.1156S} Seitenzahl, I.~R., Ciaraldi-Schoolmann, F., R{\"o}pke, F.~K., et al.\ 2013, \mnras, 429, 1156
\bibitem[Shen et al.(2010)]{2010MNRAS.407.1581S} Shen, S., Wadsley, J., \& Stinson, G.\ 2010, \mnras, 407, 1581
\bibitem[Smagorinsky(1963)]{1963MWRv...91...99S} Smagorinsky, J.\ 1963, Monthly Weather Review, 91, 99 
\bibitem[Suda et al.(2014)]{2014MmSAI..85..600S} Suda, T., Hidaka, J., Ishigaki, M., et al.\ 2014, \memsai, 85, 600
\bibitem[Suda et al.(2008)]{2008PASJ...60.1159S} Suda, T., Katsuta, Y., Yamada, S., et al.\ 2008, \pasj, 60, 1159
\bibitem[Suda et al.(2011)]{2011MNRAS.412..843S} Suda, T., Yamada, S., Katsuta, Y., et al.\ 2011, \mnras, 412, 843
\bibitem[Totani et al.(2008)]{2008PASJ...60.1327T} Totani, T., Morokuma, T., Oda, T., Doi, M., \& Yasuda, N.\ 2008, \pasj, 60, 1327
\bibitem[Tsujimoto \& Shigeyama(2014)]{2014A&A...565L...5T} Tsujimoto, T., \& Shigeyama, T.\ 2014, \aap, 565, L5
\bibitem[Tsujimoto et al.(2017)]{2017ApJ...835L...3T} Tsujimoto, T., Yokoyama, T., \& Bekki, K.\ 2017, \apjl, 835, L3
\bibitem[Wanajo et al.(2014)]{2014ApJ...789L..39W} Wanajo, S., Sekiguchi, Y., Nishimura, N., et al.\ 2014, \apjl, 789, L39
\bibitem[Williamson et al.(2016)]{2016ApJ...822...91W} Williamson, D., Martel, H., \& Kawata, D.\ 2016, \apj, 822, 91
\bibitem[Yamada et al.(2013)]{2013MNRAS.436.1362Y} Yamada, S., Suda, T., Komiya, Y., Aoki, W., \& Fujimoto, M.~Y.\ 2013, \mnras, 436, 1362
\end{thebibliography}
\listofchanges
\end{document}